\documentclass{PoS}

\title{Associated $b$- Quark Higgs Boson Production at the LHC}

\ShortTitle{b$bH$ Production}

\author{\speaker{S.~ Dawson}\\
        Brookhaven National Laboratory, Upton, New York~~ 11973\\
        E-mail: \email{dawson@bnl.gov}}

\author{C.~B. Jackson\\
        University of Texas, Arlington, Texas~~76019\\
        E-mail: \email{cbjackson@uta.edu}}
\author{P.~Jaiswal\\
        YITP, Stony Brook University, Stony Brook, New York~~11794\\
        E-mail: \email{pjaiswal@quark.phy.bnl.gov}}

\abstract{The associated production of a Higgs boson with a $b$ quark is a discovery mode for an MSSM Higgs boson at large $\tan\beta$.  We present updates on
the production rate at the LHC, along with a discussion of the importance
of the SQCD corrections from squark and gluino loops.  We also discuss
the  purely electroweak contributions.}

\FullConference{35th International Conference of High Energy Physics - ICHEP2010,\\
		July 22-28, 2010\\
		Paris France}

\begin{document}

\section{Introduction}

The search for a Higgs boson is 
one of the most important tasks for the CERN
Large Hadron Collider.  In the Standard Model, there is
a single neutral Higgs boson.
In the minimal supersymmetric model (MSSM), however, there
are five Higgs bosons--two neutral Higgs bosons, 
$h^0$ and $H^0$, a pseudoscalar, $A^0$,
and two charged Higgs bosons, $H^\pm$, and the
strategy for discovery is quite different from in the Standard 
Model\footnote{The MSSM neutral Higgs bosons are generically
denoted by $\phi$.}.  In 
the MSSM, the couplings of the Higgs bosons to  $b$ quarks can be significantly
enhanced and, for a large range of parameter space, Higgs production in association
with $b$ quarks is the major 
discovery channel. 
The hadronic production rate for the associated
 production of a Higgs boson and a $b$ quark is well 
understood \cite{Dawson:2005vi,Dawson:2004sh,Dawson:2003kb,Campbell:2004pu,Dittmaier:2003ej}. 
In a $5$- flavor number PDF scheme, the lowest order process for producing
a Higgs boson in association with $b$ quarks is $b{\overline b}
\rightarrow \phi$ when no $b$ quarks are tagged in the final state and 
 $b g\rightarrow b \phi$ when a single $b$ quark 
is tagged.  The Tevatron experiments have produced limits on both 
of these processes as a function of the $b {\overline b} \phi$  coupling.

In this note, we present results for the process
 $b g\rightarrow b \phi$ at the LHC, 
including NLO QCD corrections, NLO SQCD corrections from
squark and gluino loops\cite{Dawson:2007ur},  and the purely electroweak 
corrections\cite{Dawson:2010yz}.

\section{Basic Setup}

\begin{figure}[t]
\begin{center}
\includegraphics[scale=0.8]{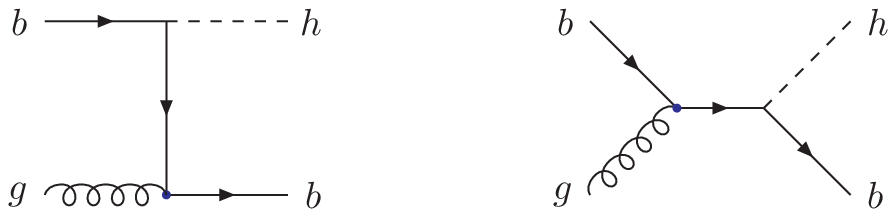} 
\caption[]{Feynman diagrams for $ g+b \rightarrow
b+ \phi$, where $\phi=h^0, H^0$ or $A^0$.}
\label{fg:bghb_feyn}
\end{center}
\end{figure}

\begin{figure}[t]
\begin{center}
\vskip 3in
\includegraphics[scale=0.4,angle=-90]{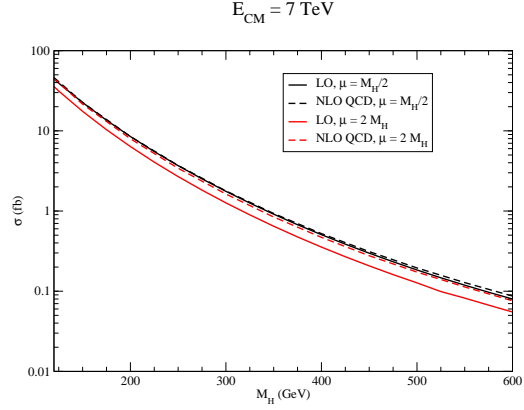}
\caption[]{
Lowest order (LO) and NLO QCD 
results for  the Standard Model process
$p p \rightarrow b ({\overline b})
H X$ at the 
LHC  with $\sqrt{s}=7~TeV$, $p_{T}^b>25~GeV$,  $\mid
\eta_b\mid < 2.5$, and $\Delta R > .4$.
The renormalization/factorization
scales are set equal to $\mu$.}
\label{fg:tev7}
\end{center}
\end{figure}

The tree level diagrams for $g + b \to b + \phi$
 are shown 
in Fig. \ref{fg:bghb_feyn}.
The lowest order (LO) rate and the NLO QCD corrected rate
 for the production of a Standard Model Higgs boson
at the LHC for the process $bg\rightarrow b H$ are shown
in Fig.  \ref{fg:tev7} for $\sqrt{s}=7~ TeV$.  The inclusion of 
the NLO QCD
corrections increases the rate significantly and also reduces the 
renormalization and factorization scale dependence.
The $b{\overline b}H$ coupling, 
$g_{b {\overline b}H}^{SM}$, is ${\overline{m_b}}(\mu)/v$, where
$v=246~GeV$ and ${\overline{m_b}}(\mu)$ 
is the ${\overline{MS}}$ running $b$ quark mass at
$1$-loop for the lowest order predictions, and at $2$-loops for 
the NLO predictions.

In the MSSM, the bottom quark Yukawa couplings to the 
neutral Higgs bosons are:
\begin{equation}
g_{b\bar{b}h^0}^{ MSSM} =  
-\displaystyle{\frac{\sin\alpha}{\cos \beta}}g_{b\bar{b}H}^{ SM}
\qquad \qquad \qquad
g_{b\bar{b}H^0}^{ MSSM}  =  
\displaystyle{\frac{\cos\alpha}{\cos\beta}}g_{b\bar{b}H}^{ SM}\,\,\,,
\label{susycoup}
\end{equation}
where $\alpha$ is the angle which diagonalizes the neutral Higgs
mass matrix and $\tan\beta$ is the ratio of the neutral Higgs boson
vacuum expectation values.
The NLO QCD corrected
rates for production of an MSSM neutral Higgs boson can be derived 
from Fig. \ref{fg:tev7} by rescaling the couplings using Eq. \ref{susycoup}.
For large $\tan\beta$, the MSSM rates can be orders of magnitude larger
than the Standard Model rates.

\begin{figure}[t]
\begin{center}
\includegraphics[bb=8 24 700 700,scale=0.32]
{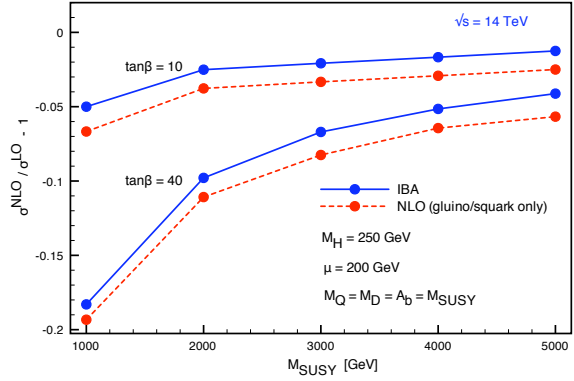} 
\vspace*{8pt}
\caption[]{
Comparison of the full SQCD calculation including
the effects of squark and gluino loops exactly
 with the approximate result (Eq. \ref{iba})
for the rate for $p p
\rightarrow b H^0$ at the LHC.  The outgoing $b$ quark 
satisfies $p_T^b>20~GeV$
and $\mid \eta_b \mid <2.5$. } 
\label{fg:MHsusy}
\end{center}
\end{figure}

The dominant SQCD radiative corrections for large squark
and gluino masses can
be taken into account by including the squark
and gluino contributions to the $b\bar
b\phi$ vertices only, i.e. by replacing the tree level Yukawa couplings by
the radiative corrected ones\cite{Hall:1993gn,Carena:1999py}. 
\begin{eqnarray}
\label{eq:bbh_rad}
g_{b\bar{b}h^0}^{MSSM} &=& 
-g_{b\bar{b}H}^{SM}\frac{1}{1+\Delta_b}
\left[\frac{\sin\alpha}{\cos\beta}-\Delta_b\frac{\cos\alpha}{\sin\beta}\right]\,\,\,,
\nonumber\\
g_{b\bar{b}H^0}^{MSSM} &=& \,\,\,\,g_{b\bar{b}H}^{SM}\frac{1}{1+\Delta_b} 
\left[\frac{\cos\alpha}{\cos\beta}+\Delta_b \frac{\sin\alpha}{\sin\beta}\right]\,\,\,,
\label{iba}
\end{eqnarray}
\noindent with
\begin{eqnarray}
\Delta_b &=& \mu\tan\beta \frac{2 \alpha_s}{3\pi } M_{\tilde g} 
I(m_{\tilde b_1},m_{\tilde b_2},m_{\tilde g})
\,\,\,,
\end{eqnarray}
where $m_{\tilde b_{1,2}}$ and 
$M_{\tilde g}$ denote the sbottom and gluino masses,
$\mu$ is the Higgs mixing parameter,
 and the  function $I$ is,
\begin{eqnarray}
\label{eq:i_function}
I(a,b,c)&=&\frac{[a^2 b^2 \ln(\frac{a^2}{b^2})+b^2 c^2 \ln(\frac{b^2}{c^2})+
a^2 c^2 \ln(\frac{c^2}{a^2})]}{(a^2-b^2)(b^2-c^2)(a^2-c^2)}\,\,\,.
\end{eqnarray}
Rescaling the $bb\phi$ couplings by Eq. \ref{iba}
 yields the result for MSSM Higgs bosons labelled
``IBA'' in Fig. \ref{fg:MHsusy}.  In Ref. \cite{Dawson:2007ur}, 
a complete calculation of
all of the contributions from squark and gluino loops to the process
$pp\rightarrow b \phi$  was presented and is shown in 
Fig. \ref{fg:MHsusy}
 as the 
curve labelled ``NLO (gluino/squark only)''.  
We see that the rescaling of Eq. \ref{iba} is an accurate approximation
for the contribution of squark and gluino loops for SUSY
masses on the TeV scale.  These contributions
can be numerically significant.  
For squark and gluino masses of ${\cal O}(1 TeV)$
and $\tan \beta =40$, they change the rate by roughly $20\%$.  
(Since the rate
is directly proportional to $\mu$, whether it is an increase or decrease 
depends on the sign of $\mu$.)

\section{Electroweak Corrections}

The $1-$loop weak corrections to the process $bg\rightarrow
b H$ consist of self energy, 
vertex, and box diagrams
containing Standard Model particles.  The
tree level process vanishes when $m_b=0$, but there are one-loop
weak contributions which do not vanish in this limit.
The electroweak contributions to the $b {\overline b}\rightarrow \phi$
MSSM
process have been computed in Ref. \cite{Dittmaier:2006cz} 
and are well approximated
by an on-shell rescaling of the $b {\overline b}\phi$ coupling.

\begin{figure}[t]
\begin{center}
\includegraphics[scale=0.35]{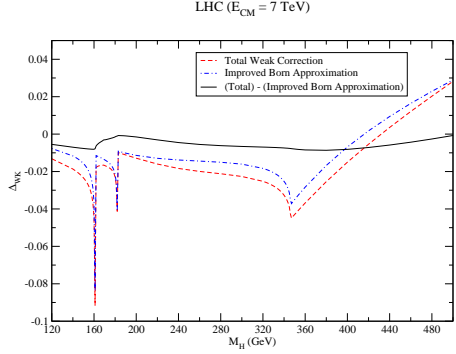}
\caption[]{
LHC results for the weak
corrections to  $p p \rightarrow b ({\overline b})
H $ with $\sqrt{s}=7~TeV$, $p_{T}^b>25~GeV$, and $\mid
\eta_b\mid < 2.5$. The solid black curve represents the contributions which
cannot be factorized into an effective ${\overline b} b H$ vertex contribution and is  less than $1\%$ for $M_H < 500~GeV$.}
\label{fg:tev7_ew}
\end{center}
\end{figure}

The cross section can be expressed as,
\begin{equation}
\sigma(bg\rightarrow b\phi)_{NLO}=\sigma(bg\rightarrow b\phi)_0\biggl(
1+\Delta_{QCD}+\Delta_{QED}+\Delta_{WK}\biggr)\, ,
\label{cordef}
\end{equation}
where $\sigma_0$ is the Born cross section derived
using the effective vertices of Eq. \ref{iba}.
Fig.~\ref{fg:tev7_ew} shows the weak corrections
to the Standard Model process, $pp\rightarrow bH$. 
The curve labelled ``Improved Born Approximation'' is derived by
rescaling the $b {\overline {b}} H$ vertex to include the electroweak
corrections obtained for
 the on-shell process $H\rightarrow b {\overline b}$,
while the curve labelled ``Total'' is an exact one-loop calculation
of the electroweak corrections to the process $bg\rightarrow bH$.
 The weak corrections are well approximated by
the Improved Born Approximation,  
with the remaining corrections always less than $1\%$.  
Except near the $W^+W$ and $ZZ$
resonances, 
$\Delta_{WK}$ in the Standard Model is
significantly smaller than the uncertainties from the QCD scale variation
and the PDF uncertainties for $M_H < 400~GeV$.
For heavy Higgs masses ($M_H\sim 1~TeV$), the weak
corrections are large (of ${\cal O}(20~\%)$).

\section{Conclusion}
The production rate for $pp\rightarrow b\phi$ for MSSM Higgs bosons is well
understood.  The QCD and SQCD one-loop corrections are large and 
must be included for an accurate prediction, while the
electroweak contributions are a few $\%$ for moderate Higgs masses.

\end{document}